# Stateless Code Model Checking of Information Flow Security

Elaheh Ghassabani and Mohammad Abdollahi Azgomi[*]

Trustworthy Computing Laboratory, School of Computer Engineering,
Iran University of Science and Technology, Tehran, Iran

E-mail: ghasabani@comp.iust.ac.ir, azgomi@iust.ac.ir

**Abstract**

Observational determinism is a security property that characterizes secure information flow for multithreaded programs. Most of the methods that have been used to verify observational determinism are based on either type systems or conventional model checking techniques. A conventional model checker is stateful and often verifies a system model usually constructed manually. As these methods are based on stateful model checking, they are confronted with the state space explosion problem. In order to verify and test computer programs, stateless code model checking is more appropriate than conventional techniques. It is an effective method for systematic testing of large and complicated concurrent programs, and for exploring the state space of such programs. In this paper, we propose a new method for verifying information flow security in concurrent programs. For the first time, we use stateless code model checking to verify observational determinism.

## 1. Introduction

Confidentiality is a key feature of computer applications such as Internet banking and authentication systems. Private data in such systems should be protected strictly. The key idea is that secret information should not be derivable from public data [1]. In a multi-level security system, data is processed in different security levels. In a scenario with two security levels, the low level is equal to public data and the high level is equal to private data. One security property of a program with a multi-level security policy is that information from a specified security level must not flow to a lower level [2]. The notion of information flow security is usually based on the non-interference property [3]. Non-interference is a model of the multi-level security policy, which intuitively means private data will not affect public data [4]. This property confirms that sensitive information is kept hidden while the program is running. It guarantees the initial values of the private variables have no effect on the final values of the public

---

[*] Corresponding author. Address: School of Computer Engineering, Iran University of Science and Technology, Hengam St., Resalat Sq., Tehran, Iran, Postal Code: 16846-13114, Fax: +98-21-73021480, E-mail: azgomi@iust.ac.ir.

variables [2, 5].

In the literature, it has been a perennial challenge to verify information flow security in concurrent programs. There is a lot of work on enforcing information flow security using type systems at the compiler level. Recently, it has been tried to deal with dynamic concepts of programs in type systems, but, because of the static nature, all the related work in this area has taken a restrictive approach. Static techniques have difficulty enforcing information flow security policies into concurrent programs. One of these difficulties is non-deterministic thread scheduling [6]. Information flow security depends on the scheduling policy in concurrent programs. A program which is secure under a particular scheduler may not be secure under another scheduler [6, 7].

Generally speaking, model checking [8, 9] is a promising method for detecting and debugging deep concurrency-related errors [9-11]. Most of the existing model checkers are stateful or/and model-based. We refer to these tools as conventional model checkers. In other words, conventional model checking includes stateful and model-based techniques. In several studies, such as [2, 5], model checking methods were employed to verify information flow security. However, all of them used conventional model checking techniques to verify an abstract core of a simple imperative language. In addition, most work in this area has concentrated on verifying sequential programs (such as [2, 12]). It is suited to verify information flow security via model checking [8], but the problem with this approach is that conventional model checkers are rather inappropriate to verify realistic and large programs.

A stateful model checker systematically explores the whole state space of a given system saving all the visited states, and then verifies the system by using the captured states. In order to use a model-based model checker, the user usually has to manually model the target system and give it as input to the model checker. So the validity of the verification result relies on that model, hence it is necessary that the input model conform to the target code. Users are usually expected to formally specify the system properties. Then, the model checker verifies whether the constructed model satisfies the specified properties [9].

In the late 1990s, "code model checking" or "software model checking" emerged as a new trend in the model checking field. This approach is more appropriate for the software world because it makes it possible for users to verify their systems without having to manually model them [13, 14]. Code model checkers, such as JPF [15], GMC [16], CHESS [17], and SLAM [18], verify programs code directly [11, 14]. These tools may be either stateful or stateless (however, it is possible for a code model checker to apply a combination of these methods). Code model checkers may be model-based; i.e. they try to automatically extract an abstract model from code and explore the state space of the extracted model.

Although stateful tools are powerful enough to verify sequential programs, they usually encounter the state space explosion problem during verification of parallel programs. Due to saving all the state space, when the concurrency level rises, we will face more complexity, and the state space will grow exponentially [9]. This is where stateless code model checking comes into use. This method does not need to save any program states. It is particularly suited to explore the state space of large programs, because precisely capturing all the essential states of a large program could be a daunting task [10, 19]. The global variables, heap, thread stacks, and register contexts are only parts of one state of a running program. Even if all the states of a program could be captured, processing such large states would be very expensive [10, 20].

In this paper, we propose a new method for verifying non-interference in concurrent programs. The proposed method uses stateless code model checking (instead of conventional techniques) to automatically verify *observational determinism* [7, 21, 22]. Observational determinism is a definition of non-interference for concurrent programs. The main contribution of this paper is to apply stateless code model checking to verifying observational determinism. As yet, several formal definitions have been proposed for observational determinism. However, it is still a challenge to practically verify observational determinism property despite its theoretical power. Some of the stateless model checking techniques, such as [10, 23-27], have demonstrated the power of stateless model checking to verify large and real-world programs. Seeking to find an effective verification method, we have decided to use stateless code model checking for this purpose.

One of the most important contributions of this paper is that observational determinism is checked on single traces instead of two traces as in the literature based on the technique of self-composition. Additionally, the method proposed in the paper can be used for detecting trace stutter equivalence in real-world concurrent programs, which could have useful applications.

The remainder of this paper is organized as follows. Section 2 gives some background on stateless code model checking and the formal definitions used in the paper. This section describes the concept of observational determinism as well. Section 3 presents our method for verifying observational determinism. Section 4 surveys and discusses the related work on information flow security. Finally, Section 5 mentions some concluding remarks as well as avenues for future work.

## 2. Preliminaries

This section presents the formal background for this paper. The first subsection briefly introduces stateless code model checking. The second subsection states some of the required formal definitions. Finally, the last subsection formally defines observational determinism and its related concepts.

### 2.1. Stateless Code Model Checking

Stateless model checking is a useful state space exploration technique for systematically testing of complex real-world software [10, 19, 28]. The notion of stateless model checking was proposed simultaneously with the appearance of code model checking by Godefroid in [19].

A stateless model checker explores the state space of a program without capturing any program states. The program is executed under the control of a special scheduler, which systematically enumerates all execution paths of the program obtained by the nondeterministic choices. In other words, the scheduler controls the nondeterministic execution of threads. Existing stateless model checkers are execution-based (not model-based) so they execute programs in their runtime environments (or an emulated environment of real runtime environment) and explore the program state space by concretely executing. That is, the stateless model checker explores the state space of a program by repeating the process of executing the program with different scheduling choices. [10, 28, 29].

As this method is applied to the source code level, it is very similar to software testing. In fact, it is a systematic testing method. A stateless model checker systematically explores all possible interleavings of threads in the program under specific input for that program [30]. In terms of state space exploration, many techniques have been proposed in the area of stateless model checking so far, including dynamic partial order reduction [25], fair scheduling [10], symmetry reduction [26], distributed dynamic partial order reduction [23, 31], which make it possible for stateless model checkers to explore the meaningful part of the state space. All these techniques have proved the power of stateless model checking to verify large programs.

In this paper, we take the approach proposed by Musuvathi and Qadeer [10] as fair stateless model checking. In this method, it is unexpected for a program not to terminate under a fair schedule. In other words, non-termination under fair scheduling is potentially an error. Our method is also applicable to programs that are expected to terminate under all fair schedules. However, these programs may not terminate under unfair schedules. Such programs are called fair-terminating [10].

The concept of fair-terminating programs is based on the observation of the test harnesses for real-world concurrent programs. Practically speaking, concurrent programs are combined with a suitable test harness that makes them fair-terminating when it comes to testing. By doing so, every thread in the program is eventually given a chance by the fair scheduler to make progress, which guarantees the (correct) program as a whole can make progress towards the end of the test. Such a test harness can be created even for systems such as cache-coherence protocols that are designed to "run forever"; the harness limits the number of cache requests from the external environment.

Therefore, our method is applicable to a fair stateless model checker that has an explicit scheduler that is (strongly) fair and at the same time sufficiently nondeterministic to guarantee full coverage of safety properties. Such a fair scheduler has been implemented in the CHESS model checker [10, 17, 27] as well as DSCMC [32]. In this context, fairness is defined as follows: every thread that is enabled infinitely often gets its turn infinitely often.

It should be noted that stateless model checkers expect the program under test to eventually terminate. In other words, practically, it is not possible for a stateless model checker to identify or generate an infinite execution. They have some mechanism to deal with non-terminating programs [10, 19, 23, 27, 30]. For example, they may ask the user to set a large bound on the execution depth. This bound can be orders of magnitude greater than the maximum number of steps the user expects the program to execute. The model checker stops if an execution exceeds the bound, and reports a warning to the user. The user can examine this execution to see whether it actually indicates an error. In the rare case it is not, the user simply increases the bound and runs the model checker again [10]. Above all, the stateless model checkers that do not apply the fair stateless model checking method, like Inspect [30, 33], are unable to properly verify the nonterminating programs that are fair-terminating. This is because they cannot detect existing cycles in the state space [10, 19, 30].

The description of the fair stateless model checking algorithm is lengthy and complicated so we do not state it in this paper. However, we use this notion as part of our method. But, there is no need to state it in detail. When it comes to explaining our method, "fair state space exploration" represents the use of its related algorithm elaborated in [10].

## 2.2. Program Model

Transition systems are often used in computer science as models to describe the behavior of systems. They are directed graphs where nodes represent states and edges model transitions, i.e. state changes. A state describes some information about a system at a certain moment of its behavior. A state of a sequential computer program indicates the current values of all program variables together with the current value of the program counter that indicates the next program statement to be executed. We use transition systems with atomic propositions for the states. Atomic propositions (*APs*) intuitively express simple known facts about the states of the system under consideration [9]. The definitions proposed in this section are taken from [9].

**Definition 1.** A transition system *TS* is a tuple *(S, Act, →, I, AP, L)* where *S* is a set of states, *Act* is a set of actions, $\rightarrow \subseteq S \times Act \times S$ is a transition relation, $I \subseteq S$ is a set of initial states, *AP* is a set of atomic propositions, and $L : S \rightarrow 2^{AP}$ is a labeling function.

Here, $2^{AP}$ denotes the power set of *AP*. For convenience, we write $s \xrightarrow{\alpha} s'$ instead of *(s, α, s′)* ∈ →. The intuitive behavior of a transition system can be described as follows. The transition system starts in some initial state $s_0 \in I$ and evolves according to the transition relation →. That is, if *s* is the current state, then a transition $s \xrightarrow{\alpha} s'$ originating from *s* is selected non-deterministically and taken, i.e. the action α is performed and the transition system evolves from state *s* into the state *s'*. This selection procedure is repeated in state *s'* and finishes once a state is encountered that has no outgoing transitions.

For a sequential program, a program graph (*PG*) over a set of typed variables is a digraph whose edges are labeled with conditions on these variables and actions. Intuitively, a program graph is like the program control flow graph (CFG), in which executing an instruction changes the program state. Obviously, each sequential program has a program graph, which can be interpreted as a transition system [9]. After interpretation, states of the transition system are pairs of the form *(l, η)* where *l* is a program location and *η* denotes values of all the program variables in location *l*. For program *P*, the transition system of the program graph is denoted by *TS (PG)*.

A concurrent system is composed of a finite set of threads or processes whose state space is defined by using dynamic semantics in the style of transition systems. Each process executes a sequence of statements in a deterministic sequential programming language. Threads are a particular type of processes that share the same heap [25]. A multithreaded program can be modeled as a concurrent system, which consists of a finite set of threads, and a set of shared objects. Threads communicate with one another only through shared objects [26].

The transition system of a multithreaded program with *n* threads running in concurrent is defined as *TS (PG_1 ||| PG_2 ||| ... ||| PG_n)* where $PG_i$ is the program graph of $i^{th}$ thread and ||| denotes the interleaving operator. Interleaving means the nondeterministic choice between activities of the simultaneously acting threads.

**Definition 2.** Path: Let $\pi = s_0 s_1 s_2 ... s_n$ be a finite path of transition system *TS*, where $s_i$ is a state of the transition system. A path is formed from a sequence of actions. Thus, $\pi$ is formed from the execution of actions $\alpha_j$ for *j = 0, 1, 2, ...,n* such that $s_0 \xrightarrow{\alpha_0} s_1 \xrightarrow{\alpha_1} s_2 ... \xrightarrow{\alpha_n} s_n$.

Based on the explanation as to stateless model checking, we consider all paths of *TS ($PG_p$)* as finite.

**Definition 3.** Trace: We consider sequences of the form $trace(\pi) = L(s_0) L(s_1) L(s_2) ... L(s_n)$ as traces of the transition system which register the (set of ) atomic propositions that are valid along the execution. $L(s_i)$ is a function that returns a subset of atomic propositions of $s_i$.

A path of *s* is a path started from *s*. *Paths(s)* is called the set of all the paths of *s*. *Traces(s)* is the set of traces of *s*. It is supposed that *Traces(s) = trace( Paths(s) )*, where *trace(π)* is a function that maps path *π* to its corresponding trace. In the same way, *Paths(TS)* is a set of all the paths in *TS*, from which the set of traces in *TS*, *Traces(TS)*, is obtained.

When a program is executed, the execution is equal to a path of its state space (i.e. a path of *TS(PG)*). The previous section introduced the idea of stateless model checking. The algorithm of state space exploration in stateless model checking is precisely described in [19, 29]. Intuitively, a stateless model checker explores the state space of a program by concretely and continuously re-executing the program such that the model checker generates a different thread scheduling scenario for each execution. We do not intend to elaborate the algorithms used in stateless model checking [23, 25, 27, 29] but it should be pointed out that the process of stateless model checking contains finite *iterations*. Each iteration is equivalent to the execution of program *P* under the control of its scheduler.

**Definition 4.** Each execution of program *P* (i.e. each iteration of stateless model checking for *P*) is a path in the transition system of the program graph of *P* (i.e. a path in $TS(PG_P)$).

**Definition 5.** Stutter-equivalence of finite paths: let $TS_i = (S_i, Act_i, \rightarrow_i, I_i, AP, L_i)$ be transition systems, $Paths(TS_i)$ is the set of finite paths in $TS_i$, $i = 1, 2$, and $\pi_i \in Paths(TS_i)$. Finite paths $\pi_1$ in $TS_1$ and $\pi_2$ in $TS_2$ are stutter equivalent, denoted $\pi_1 \triangleq \pi_2$, if there exists a finite sequence $A_0 \ldots A_n \in (2^{AP})^+$ such that $trace(\pi_1)$ and $trace(\pi_2)$ are contained in the language given by the regular expression $A_0^+ A_1^+ \ldots A_n^+$.

Here, the notation "$^+$" matches the preceding element one or more times. The notion of stutter equivalence can be adapted to traces over the alphabet $2^{AP}$ in the obvious way.

**Definition 6.** Stutter-equivalence of traces: traces $\sigma_1$ and $\sigma_2$ over $2^{AP}$ are stutter equivalent, denoted $\sigma_1 \triangleq \sigma_2$, if they are both of the form $A_0^+ A_1^+ \ldots A_n^+$ for $A_0 \ldots A_n \in (2^{AP})^+$.

## 2.3. Observational Determinism

Observational determinism is a generalization of non-interference that can be applied to multithreaded programs [22, 34]. As the concurrent programming languages are naturally nondeterministic, it is problematic to generalize non-interference to these languages; the language semantics does not specify the order of execution of concurrent threads. Even though the non-determinism permits a variety of thread scheduler implementations, it also leads to *refinement attacks*, in which information is leaked through resolution of either nondeterministic choices or scheduler choices [22]. The refinement attacks often exploit timing flows, covert channels that have been difficult to control [35]. Elimination of these attacks is possible by extending the non-interference with observational determinism [22]. This method avoids some of the restrictiveness imposed on security-typed concurrent languages [22, 34, 36].

Intuitively, observational determinism expresses that a multithreaded program is secure when its publicly observable traces are independent of both its confidential data and its scheduling policy [22]. For observational determinism, several formal definitions have been proposed since 2003. Each definition has tried to improve the previous ones. This history is surveyed in [6, 37]. This paper is based on the definition of observational determinism proposed by Huisman *et al.* in 2011. Therefore, this section states the definition that we use in our method.

In a program with a multi-level security policy, it is generally assumed that there is a lattice $Ł$ of security labels. The lattice elements describe restrictions on the propagation of information they label; the labels higher in the lattice describe data whose use is more restricted. For the sake of simplicity, we consider a simple two-point security lattice, where the data is divided into two disjoint subsets $H$ and $L$, containing the variables with *high* (i.e. *private*) and *low* (i.e. *public*) security level respectively.

A *store* is the current state of a program memory mapping a value to the location of each program variable. Two stores $s_1$ and $s_2$ are equivalent if the values of all the variables in $s_1$ and $s_2$ are the same [6].

**Definition 7.** Stores $s_1$ and $s_2$ are low-equivalent, denoted $s_1 =_L s_2$ iff the values of all variables in $L$ in $s_1$ and $s_2$ are the same; i.e. $s_1 =_L s_2$ *iff* $s_{1/L} = s_{2/L}$ [6].

For example, suppose you have a program with two low locations (variables) $l_1$ and $l_2$. The low store for this program is a tuple of the form $(l_1.val, l_2.val)$ where $l_i.val$ denotes the value of variable $l_i$. Let the initial values for $l_1$ and $l_2$ be 0 then the initial low store is (0, 0). The low store changes during the program execution. For example, suppose $l_1$ changes to 1, thereafter the low store changes to (1, 0), then if $l_1$ changes to -1, the low store goes to state (-1, 0). Finally, when $l_2$ changes to 2, the low store also changes to (-1, 2). Consequently, the low store trace in this example is a sequence of the form "(0, 0) (1, 0) (-1, 0) (-1, 2)".

**Definition 8.** The low store trace for trace $\sigma$ is denoted by $\sigma_{/L}$, which means we only consider low store values as the elements of *AP*.

**Definition 9.** Given any two initial low equivalent stores, $s_1 =_L s_2$, program $P$ is observationally deterministic *iff* any two low store traces of $s_1$ and $s_2$ are stutter equivalent, formally: $\forall \sigma_1, \sigma_2, s_1 =_L s_2, \sigma_1 \in Traces(s_1), \sigma_2 \in Traces(s_2) \Rightarrow \sigma_{1/L} \triangleq \sigma_{2/L}$ [37].

**Theorem 1.** For any two initial low equivalent stores, if any two low store traces obtained from the execution of a program under a nondeterministic scheduler are stutter equivalent, this program is secure under any scheduling policy [37].[*]

**Corollary 1.** By straightforward induction on Definition 9 and Theorem 1, program $P$ is secure under any scheduling policy if it is observationally deterministic.

---

[*] The proof is available on [37].

## 3. The Proposed Method

This section elaborates how observational determinism can be verified with stateless code model checking. Our definition of observational determinism is formally stated in Section 2. This section proposes a new method for verifying this definition. First, we describe the main idea, and prove its soundness; next, the algorithm of the method is proposed. Then, an example is described to illustrate the algorithm. Finally, we discuss the soundness of the algorithm.

### 3.1. Method

All possible interleavings of threads (or processes) should be systematically explored by a model checker so model checking is normally free of any specified scheduling policy. That is, the scheduler of a stateless model checker attempts to generate enough schedules to achieve full state coverage. Therefore, it is appropriate to use Huisman's *scheduler-independent* definition (Definition 9) for observational determinism. Scheduler-independence means that if a program is secure under a security specification then an attacker cannot infer any secret information from it, regardless of which scheduler is used [6, 37].

Hereafter, we define some notations:

1. $SMC(P, s_0)$ is a function of stateless model checking (i.e. fair state space exploration) for program $P$ from initial states that are low equivalent to $s_0$ such that $SMC(P, s_0) \hookrightarrow Paths(s_0)$, where $\hookrightarrow$ is the notation of mapping in a function, and $Paths(s_0) \in Paths\,(TS(PG_p))$. That is to say, $SMC(P, s_0)$ generates all paths of $s_0$.
2. $TS = (S, Act, \rightarrow, I, AP, L)$ is the transition system of the program graph for program $P$ (i.e. $TS = TS(PG_p)$).
3. $AP$ is considered as a set of all possible values of the low store.
4. $i_{j,s_0}$ denotes $j^{th}$ iteration of the stateless model checking process, which denotes the process is started from $s_0 \in I$.
5. $i_{j,s_0} \hookrightarrow \pi_j$, $\pi_j \in Paths(s_0)$, $Paths(s_0) \in Paths(TS(PG_P))$. That is, $i_{j,s_0}$ leads to the exploration of $\pi_j$.
6. $P^*$ indicates that program $P$ is secure under any scheduling policy.

The proposed method assumes that the number of threads and initial states are finite. To verify (rather than test) observational determinism we use a new definition of stateless model checking, called *complete stateless model checking (CSMC)*.

**Definition 10.** *CSMC*: Let the initial low equivalent stores of program $P$ be divided into $n$ finite categories, $n \geq 1$. CSMC for $P$, denoted $CSMC_{P,\,n}$, is defined as a set of stateless model checking functions: $CSMC_{P,\,n} = \bigcup SMC\,(P,\,s_i)$ for $i = 1, 2, ..., n$.

**Theorem 2.** low store traces of program $P$ can be obtained from $SMC(P, s_0)$, denoted $\forall i_{j,s_0} \hookrightarrow \sigma_{j|L}$, where $\sigma_{j|L} \in Traces(s_0)$.

*Proof.* According to Section 2, each computer program, such as $P$, has a $PG$, from which $TS(PG_P)$ can be obtained. From Definition 3, each path can be mapped to a trace registering the atomic propositions that are valid along the execution. Definition 8 implies that $\sigma_{|L} = \sigma$ when $AP$ is defined over the possible values of the low store. Consequently, $i_{j,s_0} \hookrightarrow \pi$, $\pi \in Paths(s_0) \in Paths(TS)$, from which a low store trace can be extracted. Considering the definition of $SMC(P, s_0)$ and $Traces(s_0)$, all low store traces obtained from the iterations belong to $Traces(s_0)$:

$\forall i_{j,s_0} \hookrightarrow \sigma_{j|L}$, $\sigma_{j|L} \in Traces(s_0)$, $Traces(s_0) \in Traces(TS(PG_P))$. □

According to Definition 9, if every pair of low store traces of initial low equivalent stores is stutter equivalent, P is secure. $SMC(P, s_0)$ considers only one sort of initial low equivalent stores, whereas we need to consider the whole initial low equivalent stores in order to *verify* observational determinism. As the definition of *CSMC* is aware of the condition, it explores traces of different sorts of initial low equivalent stores.

**Corollary 2.** Program $P$ is secure under any scheduling policy if *for n = 1, 2, ..., k, and $k \geq 1$*, $\forall SMC(P, s_n) \in CSMC_{P, k}$, $\forall \sigma_{m'|L}, \sigma_{m|L} \in Traces(s_n) \Rightarrow \sigma_{m'|L} \triangleq \sigma_{m|L}$.

**Theorem 3.** Program $P$ is secure under any scheduling policy if each low store trace obtained from $SMC(P, s_x)$, $1 \leq x \leq k$, conforms to the same regular expression, where $k$ is the size of set $CSMC_{P, k}$:

$\forall \sigma_{m'|L}, \sigma_{m|L} \in Traces(s_x)$, $A_0 ... A_n \in (2^{AP})^+ \sigma_{m'|L}, \sigma_{m|L} \in A_0^+ A_1^+ ... A_n^+ \Rightarrow P^*$

*Proof.* According to Corollary 1, $P$ is secure under any scheduling policy if it is observationally deterministic. From Definition 9, a program is observationally deterministic *iff* the low store traces obtained from initial low equivalent stores are stutter equivalent. Following Theorem 2, $SMC(P, s_0)$ generates low store traces which belong to $Traces(s_0)$. Based on Definition 6, we can detect stutter equivalence of traces with a regular expression. Therefore, low store traces are stutter equivalent if they conform to the same regular expression. So based on Corollary 2, $P$ is secure if each low store trace obtained from $SMC(P, s_x)$, $1 \leq x \leq k$, conforms to the same regular expression.

$\forall \sigma_{m'|L}, \sigma_{m|L} \in Traces(s_x)$, $A_0 ... A_n \in (2^{AP})^+$, $\sigma_{m'|L}, \sigma_{m|L} \in A_0^+ A_1^+ ... A_n^+ \Rightarrow P^*$ □

We call the regular expression related to $SMC(P, s_x)$ the pattern of secure low store traces (or *secure pattern*). Theorem 3 is used to verify observational determinism. Note that for $\forall m,n$, $1 \leq m,n \leq k$, the secure pattern for the low store traces obtained from *SMC(P, $s_m$)* can be different from the pattern for *SMC(P, $s_n$)*. In fact, Theorem 3 states all traces obtained from *SMC(P, $s_x$), $1 \leq x \leq k$*, must be stutter equivalent (conform to the same regular expression) if program *P* is secure.

### 3.2. Algorithm

According to the previous subsection, if all low store traces generated in *SMC(P, $s_n$)* are stutter equivalent, the program in that part of state space pertaining to *SMC(P, $s_n$)* is secure. If there is at least one low store trace which is not stutter equivalent with others, that low store trace does not conform to the secure pattern and the program violates the security policy.

Let us illustrate the forgoing with a simple example; consider a program, *P*, with two low locations *l1* and *l2*, that are Boolean with initial values of 0. Suppose we are performing *SMC(P, $s_n$)*; at runtime, whenever either *l1* or *l2* is defined, its new value after definition is immediately monitored by the model checker. If the new value of the location recently redefined differs from its previous value, a state change of the *low location* has occurred. Suppose the program is executed such that a sequence of the form "(0, 0) (0, 1) (1, 1)" is obtained as its low store trace; this sequence denotes that the program was executed such that first, *l2* changed to 1, next, *l1* changed to 1. According to Theorem 3, all the low store traces obtained from *SMC(P, $s_n$)* are expected to be contained in the language given by the regular expression "$(0, 0)^+ (0, 1)^+ (1, 1)^+$" if the program is secure.

We describe an algorithm for detecting observational determinism, which is used in *SMC(P, $s_n$)* $1 \leq n \leq k$. For *SMC(P, $s_n$)*, after ending each iteration of stateless model checking, the model checker should run another iteration. After ending *SMC(P, $s_n$)*, the model checker also should run *SMC(P, $s_{n+1}$)*. Fig. 1 shows the algorithm of *$CSMC_{P,k}$*. In this algorithm, variable *end_of_prog* denotes whether the current iteration of stateless model checking ends or not. At the end of each iteration, this variable will be set. Variable *result* shows the result of verification.

In order to ensure information flow security, a pattern of low store traces (a regular expression) is made in $i_{1,s_n}$, and all the traces obtained from *SMC(P, $s_n$)* are expected to match the pattern. We can dynamically check whether the trace of the current execution is stutter equivalent to other traces obtained from *SMC(P, $s_n$)* or not. Our algorithm is based on state changes of the low store in the current iteration.

The process of verifying observational determinism in each *SMC(P, $s_n$)* is divided into two phases: (1) detecting the pattern and saving it as a retrievable signature (2) checking whether the low store traces are stuttering equivalent (i.e. conform to the signature). This section presents the pseudo-code of the algorithms used in each phase. The algorithm for the first phase is shown in Fig. 2. This algorithm is used in $i_{1,s_n}$ of stateless model checking. Fig. 3 shows the algorithm used in the second phase, used for $\forall i_{i,s_n}$, $i \geq 2$. Therefore, the main body of this algorithm (lines 3-19 of Fig. 3) is repeated for each $i_{i,s_n}$, $i \geq 2$.

In these algorithms, DETS [38] is a kind of hash table whereby data records are saved on the disk, which is an efficient storage mechanism. Variable *lssc* is a counter that shows the number of low store state changes. It is obvious that this number for the above example should be 3. We use this counter for saving the pattern as a retrievable concise signature. Moreover, with the help of this counter, traces stutter equivalence is detected dynamically.

```
1.  Let P have k sorts of initial low equivalent stores: s₀,…, sₖ
2.  Let Path(sᵢ) be the set of paths of sᵢ
3.  boolean end_of_prog = false   // Shared variable
4.  boolean result = false        // Shared variable
5.  sem result_IFS = 0            // Binary semaphore

6.   boolean CSMC(P, s₀, …, sₖ)
7.       boolean result_smc
8.       For i = 1 to i = k
9.            result_smc = SMC(P, sᵢ)
10.           If ! result_smc
                 // A security violation was detected
11.               Return false           // The end of CSMC_{P,k}
12.           End-if
13.      End-for
14.  Return true         // P is secure ---- The end of CSMC_{P,k}

15. boolean SMC(P, sᵢ)
16.     Spawn IFS_first_itr ( )
17.     Spawn Fair_state_exp (P, sᵢ)
18.     Wait result_IFS
19.     If ! result
20.          Kill Fair_state_exp
21.          Return false              // A security violation was detected
22.     End-if
23. Return true

24. void Fair_state_exp (P, sᵢ)
        /* Begin fair state space exploration
         * It continues until all paths of Path(sᵢ) have been explored
         */
25.     While Path(sᵢ) ≠ ∅
26.          Execute P under the fair scheduler
27.          Explore π      // π ∈ Path(sᵢ): π is the path obtained from the current execution of P
28.          When P comes to an end  // Current execution of P ends: the end of the current iteration
29.              end_of_prog = true
30.              Path(sᵢ) = Path(sᵢ) - π   // Remove π from Path(sᵢ)
31.          End-when
32.     End-while
33. Return
```

**Fig. 1**   The algorithm of $CSMC_{P,k}$

When a low location changes, we use data structure *LS* to save its information on the DETS table, where *LS.id* is the identifier of the low location that has caused the low store state to change, *LS.val* is the new value of that low location, and *LS.num* denotes the current value of *lssc*. At the implementation level, there should be a mechanism for detecting the memory locations of the variables; here, we assume the variables are distinguished by *identifiers* (i.e. *LS.id*). Also, aliasing of low variables may need to be dynamically detected at implementation.

At first, the stateless model checker starts the process of model checking with function *CSMC*, line 6 of Fig. 1. This function calls function *SMC* for each $s_n$. Function *SMC* creates two other processes in lines 16-17; one of them executes the fair stateless model checking algorithm [10] (lines 24-33) although it embodies the algorithm as a black box in lines 26-27. Another process is responsible for verifying observational determinism, which is called *ifsv*. These processes run in concurrent interacting each other.

After creating the above processes, line 18, function *SMC* waits for receiving a signal from the process verifying observational determinism. When it receives the signal for semaphore *result_IFS*, it checks the result of the verification using variable *result* on line 19. If the variable is *true* then the part of the state space related to *SMC(P, $s_n$)* is secure. Otherwise, a security violation has been detected and *SMC* returns *false* to *CSMC* at line 21. Function *CSMC* receives this return value on line 9. If it is true then *CSMC* sets up *SMC(P, $s_{n+1}$)*. Otherwise, it finds a violation and ends the process of verification (lines 10-12).

We assume that program *P* is repeatedly executed under the control of the stateless model checker to perform *SMC(P, $s_n$)*, and the tool is able to identify low locations at runtime (in each iteration). In each iteration, after initializing a low location, such as $l$ ($l \in L$), whenever the low location is redefined[*], if its new value varies from the previous one, *ifsv* is able to capture this state change of that low location (Fig. 2, line 4, and Fig. 3, line 5). When a low location changes to a new value, it causes the low store to change to a new state.

Our algorithm does not need to save all values of variables during the state change of the low store (unlike stateful model checking). As is clear from the above example, only the value of one low variable (low location) changes in each state change of the low store. Therefore, we only need to know the location that has caused the state change of the low store, as well as its new value. In other words, it suffices to store *id* and value of that location along with a number that represents how many times the low store has changed. Note that using this idea, we only store a pattern of state changes of the low store in $i_{1,s_n}$ (in order to dynamically construct the signature for other iterations after $i_{1,s_n}$). That is, the pattern is saved at $i_{1,s_n}$ by the algorithm in Fig. 2. In subsequent iterations, there is no need to store any states. It is enough to monitor the low store state at runtime, and compare it with the signature (Fig. 3).

For $i_{1,s_n}$, the instructions at lines 3-10 of Fig. 2 are executed as long as the program has not terminated. When a low location (such as *l*) changes to a new value (such as *newval*), *ifsv* catches this state change at line 4. Therefore, variable *lssc* is incremented (line 5), and then the identifier of *l*, its new value, and the current value of *lssc* are saved in the record *LS* (lines 6-8). The record is saved in the DETS table with key *lssc* at line 9. This procedure continues until the end of $i_{1,s_n}$.

After ending the first iteration, the number of state changes of the low store (i.e. *lssc*) is saved on the DETS at line 13, and now, the DETS contains the signature that low store traces should conform to. Thereafter, the model checker advances the model checking process by re-executing the program to explore another path (line 25 of Fig. 1).

```
1.   void IFS_first_itr( )
         // P is executing under the stateless model checker
2.       lssc = 0
3.       While ! end_of_prog
4.           When l ∈ L is defined to newval
5.               lssc += 1
6.               LS.id = l.id
7.               LS.val = newval
8.               LS.num = lssc
9.               Save the LS on the DETS with key LS.num
10.          End-when
11.      End-while
12.      end_of_prog = false
13.      Save the lssc on the DETS
14.      result = IFS( )
15.      Signal result_IFS
16.  Return
```

**Fig. 2** The algorithm of extracting the pattern of low store traces for *SMC(P, s$_n$)*

After $i_{1,s_n}$, the algorithm shown in Fig. 3 is used until the end of *SMC(P, s$_n$)*. The process ends when all execution paths obtained from threads interleaving have been explored. At last, function *IFS* returns either *true* or *false* to function *IFS_first_itr*, and the *IFS_first_itr* function saves its verification result into variable *result* signaling to *SMC* that it is time to terminate *SMC(P, s$_n$)* (lines 14-15 of Fig. 2). *SMC* receives the signal at line 18 of Fig. 1.

For $\forall i_{i,s_n}$, $i \geq 2$, when a low location such as *l* changes to a new value such as *newval* causes the low store to change. The state change is captured by *ifsv* (Fig. 3, line 5). Therefore, the record with key

---

[*] In compiler science, it is said that a location (or variable) is defined (or redefined) when a write access to that memory location takes place.

"*lssc* + 1" is fetched from the DETS, and saved in variable *temp* (Fig. 3, line 6), which represents the low location that caused the state change of the low store in the signature. If both identifier and value of the variable in the retrieved record (*temp.id* and *temp.val*) are equal to *id* and *newval* (line 7), the state change is accepted; otherwise, a security violation is revealed (lines 10-12). In the former case, *lssc* is incremented, and *ifsv* goes on monitoring. After ending each iteration, the current value of counter *lssc* should be checked to be equal to the number of the state changes of the low store in the signature (lines 16-19). If they are not equal, it represents a security violation. The reason for doing this is that Definition 9 does not allow low store traces to be stutter and prefix equivalent [6], hence the number of changes of the low store in all traces should be equal if the program is secure.

```
1.  boolean IFS( )
2.      Foreach iteration in SMC(P, sᵢ)
3.          lssc = 0
4.          While ! end_of_prog
5.              When l ∈ L is defined to newval
6.                  Fetch the tuple with key "lssc + 1" and Save it in temp
7.                  If temp.id == l.id && temp.val == newval
8.                      lssc += 1
9.                  End-if
10.                 Else
11.                     Return false              //A security violation occurred
12.                 End-else
13.             End-when
14.         End-while
15.         end_of_prog = false
16.         Fetch lssc from the DETS and Save it in stn
17.         If stn != lssc
18.             Return false                      //A security violation occurred
19.         End-if
            /* Else
             *     The verification process continues for another iteration
             *     Go to the beginning of the Foreach loop
             */
20.     End-foreach
21. Return true
```

**Fig. 3** The algorithm to check stutter-equivalence of traces *SMC(P, $s_n$)*

## 3.3. Discussion on Complexity

Our method needs to save only a short signature (in comparison with a trace) on the disk. The algorithm removes all redundancies during storage of the signature and pattern matching. Only in the first iteration, the signature is constructed based on state changes of individual *low locations*, instead of state changes of the *low store* (low store = all the low locations). Fig. 4 illustrates how to save the state changes of the low store. The algorithm of Fig. 3 only examines the *current* low store in the main memory without capturing the program state space. Using the signature, the regular expression expected to be the secure pattern is constructed on-the-fly and gradually as stutter equivalence is being verified dynamically. The example in section 3.3 illustrates how to do that. To sum up, as far as the main memory is concerned, no state is saved and the algorithm needs only simple variables which were described. However, as you can see, the number of the records saved on the disk increases in the number of the state changes of the low store.

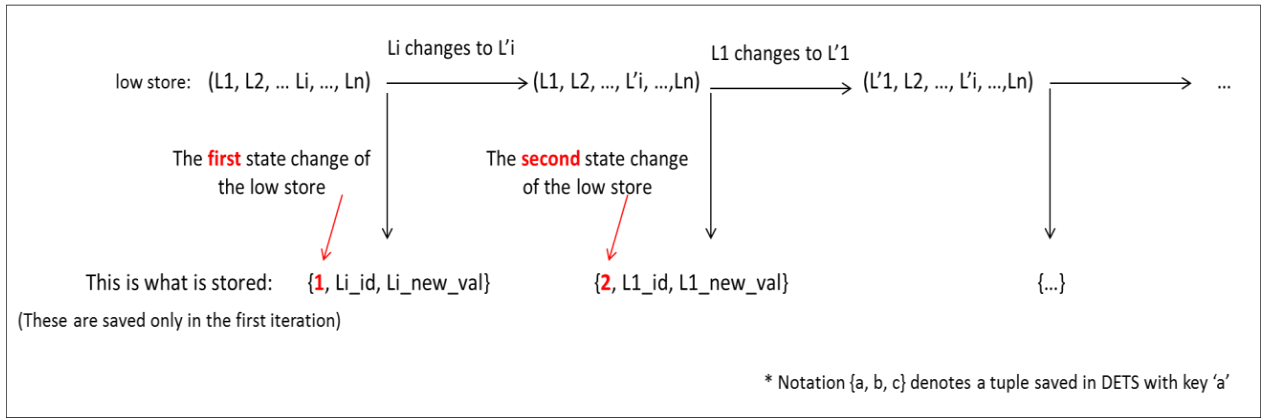

**Fig. 4**   An example of storing the signature

Please note that the algorithm in Fig. 1 stands for the same stateless model checking algorithm. This technique can be really time-consuming for complicated and large programs. One the positive side, stateless model checking does not face state space explosion and its capability to cope with real-world concurrent programs has been proved in practice. Therefore, when it comes to time complexity of the method, we consider our own algorithms in Fig. 2 and 3. As described, these algorithms are event-based; this fact is represented with the *While* loop in these algorithms. Also, the *Foreach* loop in Fig. 3 is related to iterations of stateless model checking. In fact, each iteration itself contains the *While* loop which captures value changes of low variables. Whenever the value of a low location changes, the algorithm needs some simple read and write actions. Thus, the time complexity of this algorithm is linear in the number of such changes.

The next subsection is a simple example to illustrate the algorithm. When it has been clarified how the algorithm works, we discuss its soundness.

## 3.4. Example

Consider the following program, *P*, where $h \in High$ and $l1, l2 \in Low$ (the low store of this program is a couple of the form *(l1.val, l2.val))*.

$$l1 := 0; \; l2 := 0; \; h = 0;$$

$$\{\{if \; (l1 == 1) \; then \; (l2 := h) \; else \; skip\} \; |||\; l1 := 1 \; |||\; h := 1;\}$$

This program is not secure because the value of *h* can be revealed through observing low store traces. A hacker who sees *l2* is always 0 can infer $h = 1$ when the program terminates in a state where $l2 = 1$. As $s_0 = (0, 0)$ is the initial low store, and there is only one sort of initial low equivalent stores ($s_0$ is always $(0, 0)$), so we have $CSMC_{P, 1}$. Therefore, we only need to describe $SMC(P, s_0)$.

Let $T_i$ be a thread in the above program where $T_1$ is the left-most side of the interleaving operator and $T_3$ is the right-most side of the operator. All the possible low store traces of the program, obtained from executing *P* under different scheduling policies (or it is better to say *obtained from SMC(P, $s_0$)*), are as follows:

1. *Execute $T_1$, $T_2$, $T_3$: $\sigma_{|L}$ = [(0, 0) (1, 0)]*
2. *Execute $T_1$, $T_3$, $T_2$: $\sigma_{|L}$ = [(0, 0) (1, 0)]*
3. *Execute $T_2$, $T_1$, $T_3$: $\sigma_{|L}$ = [(0, 0) (1, 0) (1, 0)]*
4. *Execute $T_3$, $T_1$, $T_2$: $\sigma_{|L}$ = [(0, 0) (0, 0) (1, 0)]*
5. *Execute $T_2$, $T_3$, $T_1$: $\sigma_{|L}$ = [(0, 0) (1, 0) (1, 1)]*
6. *Execute $T_3$, $T_2$, $T_1$: $\sigma_{|L}$ = [(0, 0) (1, 0) (1, 1)]*

As can be seen in the above *low store traces*, the traces are not stutter equivalent. Therefore, the program is not observationally deterministic and according to Corollary 1, the program is not secure. Now let us follow the algorithm for this example. Initially, function *CSMC* calls *SMC(P, $s_0$)* at line 9, Fig. 1. Therefore, it starts the process of stateless model checking; first, the variables are initialized by the program. Then, suppose the model checker first schedules $T_1$, next, $T_2$. When $T_2$ is scheduled, it changes low location *l1* into 1, which causes the low store to go from state (0, 0) to state (1, 0). Therefore, the first state change of the low store occurs. The *ifsv* catches this event at line 4 in Fig. 2. consequently, a tuple of the form *{1, 1, 1}* is inserted into the DETS (Fig.2, line 9), where from the left side, the first 1 denotes this is the first sate change of the low store (*LS.num*), the second 1 denotes the identifier of the location that caused the state change (i.e. *LS.id*; here, we supposed the identifier of *l1* is 1), and the third 1 is the new value of *id* after the state change (*LS.val*). Thereafter, the model checker schedules $T_3$, and the program ends. Through this execution, *ifsv* considers the regular expression of the low store traces to be "$(0, 0)^+ (1, 0)^+$", and the value of the counter *lssc* to be 1.

After $i_{1,s_0}$, the *IFS_first_itr* function calls *IFS* at line 14 of Fig. 2. After initialization in $i_{2,s_0}$, suppose that the model checker first schedules $T_1$, next, $T_3$, and finally, $T_2$. When $T_2$ is scheduled, $l1$ changes to a new value ($l1:= 1$). This situation shows the first state change of the low store occurred (i.e. *lssc* changes to 1). The *ifsv* catches the event at line 5 of Fig. 3. It should examine whether the state change conforms to the signature or not. For this reason, it fetches the record of the first low store state change from the DETS (i.e. the record with key 1) at line 6, then checks that both identifiers and values of the variables that cause the state change in the current situation and in the signature are the same. As both conditions are true in this example, the statements on lines 7-9 are executed. Then, the program ends, and the algorithm compares the current value of *lssc* with its value in the signature (lines 16-19). Due to the equality of the two values (both values are 1), this iteration also ends successfully.

Suppose the model checker schedules threads such that the third iteration conforms to *low store trace 3*. Therefore, this iteration ends successfully too. Thereafter, $i_{4,s_0}$ ends similarly. Let $i_{5,s_0}$ correspond to *low store trace 5*. A security violation will occur when $T_1$ is scheduled. Because when $l2$ changes to 1, *ifsv* sees the number of the state changes in the current execution exceeds the number of the state changes in the signature (i.e. *lssc != stn* - line 17); therefore, an existing security violation is detected, and *IFS* returns *false* at line 18 of Fig. 3, consequently function *SMC(P,$s_0$)* also returns *false* at line 21 of Fig. 1. Then *CSMC* receives *false* at line 9 of Fig. 1 and returns *false* at line 11.

When *CSMC* returns, the process of model checking comes to an end. If its return value is *true*, it shows the program under test is secure. Otherwise, it has detected a security flaw in the program.

### 3.5. Soundness

First of all, let us introduce some notations. Let program $P$ have $n$ low locations. Therefore, its low store is a tuple like *(l0, ..., $li_1$, ..., $li_2$, ..., $li_j$,..., ln)*. In the low store, the place of location $li_k$ in relation to the place of location $li_j$ has no matter. That is, j ≤ k or j ≥ k; e.g. for locations $li_1$ and $li_2$, no matter whether the low store is *(l0, ..., $li_1$, ..., $li_2$, ..., $li_j$,..., ln)* or *(l0, ..., $li_2$, ..., $li_1$, ..., $li_j$,..., ln)*. But, for the sake of simplicity, we suppose the low store is a tuple of the form *(l0, ..., $li_1$, ..., $li_2$, ..., $li_j$,..., ln)*. However, in the following description, notations $li_j$ and $li_k$ can even be the representative of the same place. To recap, the *ifsv* algorithm exploits this fact that only one low location has changed in each state change of the low store. Theorem 4 proves the soundness of the algorithm.

**Theorem 4.** The algorithm holds two features: (1) it truly extracts the pattern; i.e. the correct regular expression can be reconstructed, and (2) it truly detects stutter equivalence of traces.

*Proof.* In regard to the first feature, we do the proof by induction. We should prove that by having tuples *(1, $li_1$, $Xi_1$) (2, $li_2$, $Xi_2$)...(n, $li_j$, $Xi_j$)*, we are able to construct the following regular expression:

$(l0, ..., li_1, ..., li_2, ..., li_j, ..., ln)^+$ $(l0, ..., Xi_1, ..., li_2, ..., li_j, ..., ln)^+$ $(l0, ..., Xi_1, ..., Xi_2, ..., li_j, ..., ln)^+$ ... $(l0, ..., Xi_1, ..., Xi_2, ..., Xi_j, ..., ln)^+$. The base case is when the low store state only changes once. We have the initial low store so the tuple *(1, li_1, Xi_1)* shows that, in the first state change, location $li_1$ changes to $Xi_1$. Therefore, we can find out that the low store after the first state change is *(l0, ..., Xi_1, ..., li_2, ..., li_j, ..., ln)*. Needless to say, low locations can be redefined, but their value may remain unchanged. Hence, a state of the low store like *(l0, ..., Xi_1, ..., li_2, ..., li_j, ..., ln)* can stay unchanged even after a (some) low location(s) has (have) been defined. Consequently, repetition, *"+"*, may occur. In the inductive step, we assume that the feature holds after *k* state changes of the low store. Therefore, we have tuples *(1, li_1, Xi_1) (2, li_2, Xi_2)...(n, li_k, Xi_k)*, and are able to construct the regular expression $(l0, ..., li_1, ..., li_2, ..., li_j, ..., ln)^+$ $(l0, ..., Xi_1, ..., li_2, ..., li_j, ..., ln)^+$ $(l0, ..., Xi_1, ..., Xi_2, ..., li_j, ..., ln)^+$ ... $(l0, ..., Xi_1, ..., Xi_2, ..., Xi_k, ..., ln)^+$. So, from this step, we know the low store is like *(l0, ..., Xi_1, ..., Xi_2, ..., Xi_k, ..., ln)* after *k* state changes. Then, we need to show that after *(k+1)* state changes, we can construct $(l0, ..., li_1, ..., li_2, ..., li_j, ..., ln)^+$ $(l0, ..., Xi_1, ..., li_2, ..., li_j, ..., ln)^+$ $(l0, ..., Xi_1, ..., Xi_2, ..., li_j, ..., ln)^+$ ... $(l0, ..., Xi_1, ..., Xi_2, ..., Xi_k, ..., ln)^+$ $(l0, ..., Xi_1, ..., Xi_2, ..., Xi_k, ..., Xi_{k+1}, ..., ln)^+$. When we have the tuple *(k+1, li_{k+1}, Xi_{k+1})* then, using key *k+1*, the algorithm detects that low store *(l0, ..., Xi_1, ..., Xi_2, ..., Xi_k, ..., ln)* has turned into *(l0, ..., Xi_1, ..., Xi_2, ..., Xi_k, ..., Xi_{k+1}, ..., ln)*. Therefore, as the algorithm saves tuples *(1, li_1, Xi_1) (2, li_2, Xi_2)... (j, li_k, Xi_k), (j, li_{k+1}, Xi_{k+1})*, it is able to exactly construct its corresponding regular expression. □

For the second feature, we should prove that, during reconstruction of the regular expression, the algorithm is able to detect whether the current trace violates the pattern. That is to say, the algorithm should detect traces stutter equivalence correctly. Suppose that the current trace is not stutter equivalent to the other traces. So it should generate a different regular expression based on Definition 6. In light of the above explanation as to the first feature, suppose the first inconsistency in the current trace appears while the algorithm is reconstructing $k^{th}$ word of the regular expression. Therefore, when *ifsv* fetches *(k, li_k, Xi_k)*, it expects $k^{th}$ state change in the current trace to be brought about when $li_k$ changed to $Xi_k$. If this has not happened then it should detect the violation. As can be seen in Fig 3, on lines 7-13 the *ifsv* algorithm is able to do so exactly.

## 4. Discussion and Related Work

There are several mechanisms for enforcing and checking information flow security policies (IFS) in computer programs. Static techniques are conventional solutions in the field. Dynamic techniques are other solutions that are more precise and costly [39]. Language-based techniques are one strong approach for enforcing and verifying IFS, whose main idea is to design and implement programming languages that would be inherently secure. For instance, JIF [40] and Jeddak [39] are two such languages, which are extensions of Java. Although this approach is advancing, it is specific to particular programming languages being inapplicable to the legacy code.

Another approach in language-based techniques is to analyze the program source code to discover security leaks [41]. But many of these analyses are imprecise, resulting in false alarms because they are not flow-sensitive, context-sensitive, or object-sensitive. However, Hummer *et al.* [41, 42] have tried to compensate for this problem. Generally, some approaches to program analysis, such as precise inter-procedural, dataflow analysis, and program slicing, are alternatives to verification techniques (e.g., model checking).

There is a lot of work based on type systems and compiler techniques in the field though type systems are so inflexible that changing their security policies is an arduous task [2, 34]. Recently, some type systems have been proposed to address dynamic issues of concurrent programs and enforce confidentiality in multithreaded programs. They often aim to prevent the disclosure of secret information brought about by the thread timing behavior of a program, i.e. secret information cannot be derived from observing the internal timing of actions. Therefore, these type systems are often very restrictive, which makes programming rather impractical [6, 22]. One of the difficulties in verifying IFS for multithreaded programs is non-deterministic thread scheduling. The fact that a program is secure under a particular scheduler does not imply that it is secure under another scheduler [6, 7].

Some work has tried to specify information flow security as a safety property so as to be able to verify that property with existing verification methods. Francesco *et al.* [2, 5] characterized IFS in computational tree logic (CTL) using an early restrictive definition of non-interference. Boudol [43] specified a safety property for IFS that it could be used in type systems. Terauchi and Aiken [44] transformed the verification of IFS to a safety problem using self-composition concepts [45]. However, most such work concentrated on sequential programs.

Type-based approach is insensitive to control flow and rejects many secure programs. Therefore, recently, self-composition [45, 46] has been advocated as a way to transform the verification of information flow properties into a standard program verification problem (such as [6, 7, 44]). A common solution to the verification problem is to use model checking techniques. Sun *et al.* [12] proposed a method for model checking information flow for an imperative language however concurrency issues were not addressed in their work either. In their method, the model was characterized with a pushdown system. They used a semantic-based approach by self-composing symbolic pushdown system and specified non-interference property with linear-temporal logic (LTL) formula. Then, the LTL-expressed property verified with the Moped model checker [47].

Most of the related work [44, 45, 48] only considered the input-output behavior of programs while defining the non-interference property, which is not appropriate for multithreaded programs. Huisman *et al.* [34] characterized the non-interference property in CTL* (an extension of CTL [9]). The characterization considered the whole execution traces and thus amended the foregoing. Although their approach is suitable for multithreaded programs, they suggested using conventional model checking and self-composition to verify the characterization [7]. However, there is not a ready-made model checker to verify CTL* properties [7, 34].

Most of the previous work using model checking to verify IFS, such as [2, 5, 7, 12, 49], applied conventional model checking; all of them used *stateful* model checking and rarely applied *code* model checking. Unno *et al.* [49] took a fairly different approach among the work. Their method is a combination of type-based analysis and model checking; suspicious execution paths, which might cause insecure information flow, were firstly found by using the result of a type-based information flow analysis, and then a model checker was employed to check if the paths were indeed unsafe. However, they also used conventional model checking and did not consider concurrency issues.

In the literature, different definitions of confidentiality have been proposed for multithreaded programs. In 1995, Roscoe [21] stated the behavior of a program that can be observed by an attacker should be deterministic. To capture this formally, Zdancewic and Myers [22] introduced the notion of *observational determinism* in 2003. They introduced observational determinism as a scheduler independent notion of security for concurrent programs [36]. In fact, they proposed the first *formal definition* of observational determinism. Some research papers, such as [6, 34, 49], tried to amend Zdancewic's definition since then. Here, we mention some existing definitions of observational determinism in the following. For a more detailed discussion, please see [6].

Given any two initial low equivalent stores, $s_1 =_L s_2$, a program *P* is *observationally deterministic*, according to:

- Huisman *et al.* [34]: *iff* any two low location traces are stutter equivalent (2006, *scheduler-independent definition*).
- Huisman *et al.* [6]: *iff* (1) any two low location traces should be stutter equivalent and (2) given a low store trace starting in store $s_1$, for any low equivalent store $s_2$ there exists a low store trace starting in $s_2$ such that these traces are stutter equivalent (2011, *scheduler-specific definition*).
- Huisman *et al.* [37]: *iff* any two low store traces are stutter equivalent (2011, *scheduler-independent definition*).

The authors used conventional model checking to verify the above properties. The main hurdle is that conventional model checkers run into state space explosion. Therefore, it is required to improve model checking algorithms for the specific need of the observational determinism property. In addition, they did not use *code model checking* techniques and suggested using *model-based* techniques.

In addition, when it comes to using model checking as a verification method, observational determinism has usually been verified by employing self-composition [45]. From the security literature it is known how to transform the problem of checking observational determinism on a single program to the problem of checking a safety property on two copies of the program running in parallel (self-composition). The idea presented in this paper is that for checking observational determinism on single traces. It suffices to compare the output of that trace to the unique sequence of outputs of the program (which could be extracted from the first program run). Thus, unlike the descriptions in [6, 7, 37, 44, 45], we do not have to consider pairs of program runs, but only single runs, which will indeed simplify the checking process significantly.

In more recent work, Ngo *et al.* [1, 50] proposed a verification method for observational determinism. In [1], authors have presented a model checking algorithm for probabilistic programs, which aims to verify the scheduler-specific definition of observational determinism. As the verification of observational determinism in [6, 37] is rather complex and cannot be handled efficiently by the existing conventional model checkers, authors in [1, 50] proposed algorithms to verify a modified version of the definition of observational determinism. Their approach is parameterized by a particular scheduling policy, thus it can be checked whether the program will be secure under a particular scheduling policy. If the scheduling policy changes, the property has to be re-established for this new scheduling policy. It thus only gives a guarantee for a particular scheduling policy. In [50], authors used symbolic model checking to verify observational determinism. Symbolic model checking is also a powerful tool for verifying concurrent programs. But, our approach is quite different while utilizing the power of stateless code model checking. Even so, it would be an interesting future work to combine both stateless model checking and symbolic model checking to verify observational determinism in large and complex programs.

It should be pointed out that there are also other approaches to defining observational determinism. For example, recently, in [51] an alternative definition for observational determinism and its verification has been proposed. The proposed verification method makes use of program dependence graphs to check observational determinism. The proposed definition of observational determinism is also analogous to the scheduler-independent definition of observational determinism. But, in [51], instead of considering low values, the trace definition is based on low operations (read or write on a low variable). In terms of time complexity, this method has a rather high time complexity ($O(n^3)$).

It is fairly inexpensive to detect stutter equivalence of traces and check observational determinism by stateless model checking (in comparison with stateful model checking). To check observational determinism, stateful model checkers need to capture all the reachable states of the program, and compare every pair of traces as well as each trace with all the other traces. For this reason, they face the state space explosion problem. From one point of view, this paper proposes an algorithm for detecting stutter equivalence of traces in stateless model checking as well. In our method, it is enough to save only one pattern of state changes of low locations, and then stutter equivalence is dynamically computed during stateless model checking iterations. By using the method, we can detect observational determinism with stateless code model checking for the first time.

As systematic path exploration in stateless model checking is time-intensive, the verification problem will probably transform to systematical testing while verifying observational determinism in large programs. However, if we perform distributed stateless model checking, we may be able to cope with the problem. Apart from being a systematic and automatic security testing, this method is more applicable than conventional model checking to the verification (testing) of observational determinism. There are some serious problems using conventional model checking, including code modeling and state space explosion.

Conventional model checkers are state-based and need to store at least part of the program state space. Even for a simple program with just a small number of variables, the state space can be rather extreme. For instance, a program with ten locations, three Boolean variables and five bounded integers (with domain in $\{0,...,9\}$) has $10 \times 2^3 \times 10^5 = 8,000,000$ states. If a single bit array of 50 bits is added to this program, for example, this bound grows even to $800,000 \times 2^{50}$! This observation clarifies why the verification of data-intensive programs (with many variables or complex domains) is extremely hard. Even if there are only a few variables in a program, the state space that must be analyzed may be very large [9]. Hence, it may be impractical to even check observational determinism with conventional model checkers. In fact, stateless model checking is a *systematic* testing suited to explore the state space of the large programs whose state space exploration is impractical using state-based methods.

## 5. Conclusions and Future Work

This paper proposes a new verification method for ensuring confidentiality in multithreaded programs. In the literature, different definitions of confidentiality have been proposed for multithreaded programs. We follow the approach advocated by Roscoe [21], which states that the behavior which can be observed by an attacker should be deterministic. Observational determinism is one of the most powerful properties for ensuring information flow security in multithreaded programs. There is plenty of work on formalizing observational determinism as a (safety) property, which aims to convert the verification of observational determinism into a standard program verification problem, and use (conventional) model checkers. But, this paper proposes a new verification method for observational determinism based on stateless code model checking.

As yet, all model checking techniques used to verify IFS (such as [2, 5, 7, 12, 49]) fall into the category of conventional techniques; As they are state-based, they suffer from the state space explosion problem. Some of conventional model checkers are model-based and expect users to manually model their systems. So the validity of the verification result relies on that model, hence it is necessary that the input model conform to the target code. However, modeling is a costly process that requires special background knowledge and skill. Besides, there is a big conceptual gap between programming and modeling languages that makes modeling real-world programs more difficult.

Stateless model checking is a time-intensive process due to running the program concretely and exploring its entire execution paths. Hence, the applicability of the proposed method is highly dependent on its implementation. In the worst case, we can use the method for security testing or debugging. However, implementing the presented method in a parallel and distributed model checker indeed plays an important part in reduction of the verification cost. As the cycles of functions *CSMC* and *SMC* (Fig. 1) are independent, it is possible to distribute them among several computational nodes so that we can reduce verification time, and increase coverage of the state space. The advantage of distributed stateless model checking over distributed stateful model checking is that the stateless approach needs much less memory management efforts. Currently, we are implementing our method in a new parallel and distributed stateless model checker, DSCMC [32].

Our method assumes that the initial values of high variables are changed in each iteration of stateless model checking; because many security violations are revealed provided that the initial values of high variables change. At the implementation level, different methods may be applied to fulfill this need. As a whole, in (stateless) model checking, precisely handling non-deterministic input may be impractical for large programs without using symbolic techniques. Then practically speaking, our method is likely to be transformed to systematically testing. But this is kind of *security testing automatically performed* with the awareness of information flow control. To remedy the problem, the method must be improved by combining with test generation techniques (such as white-box fuzz testing [52] and symbolic execution [53]) to cover more execution paths. The method proposed by Godefroid *et al.* in [52, 54] can be effective for this purpose.

## Acknowledgement

We are grateful to Iran National Science Foundation (INSF) for financial support of this research.